\documentclass[12pt]{article}
\usepackage{epsfig}

\def\e{\begin{equation}}
\def\f{\end{equation}}
\def\=#1{\overline{\overline #1}}
\def\-#1{{\bf #1}}
\def\o{\omega}
\def\E{\epsilon}
\def\E0{\epsilon_0}

\def\M{\mu}
\def\M0{\mu_0}

\def\.{\cdot}

\def\l#1{\label{eq:#1}}
\def\r#1{(\ref{eq:#1})}
\def\am{\left(\begin{array}{c}}
\def\amm{\left(\begin{array}{cc}}
\def\a{\end{array}\right)}

\title{Isotropic microwave composites with negative refraction}
\author{ C.R. Simovski$^1$ and B. Sauviac$^2$}
\date{\today}

\begin{document}
\maketitle

\noindent $^1$St.\ Petersburg Institute of Fine Mechanics and Optics,
Russia

\noindent $^2$DIOM Laboratory ({\it \textbf{D}evices and
\textbf{I}nstrumentation
   for \textbf{O}ptoelectronics and \textbf{M}icrowaves}),
Universit\'e Jean Monnet, Saint Etienne, CEDEX 42023, France

 \vskip 1cm

\section*{Abstract}

The properties of artificial isotropic microwave composites which
would possess simultaneously negative permittivity and permeability
are studied theoretically. Four kinds of composites are considered.
Two of them concern media with split-ring resonator particles with
different particle arrangements. The other two
are realized with Omega particles. An analytical antenna model
of the electromagnetic behavior of a split-ring resonator (SRR) is
suggested and verified  by numerical simulations. Next, material
parameters of composite media made with SRR or Omega particles are
calculated. It is shown that the Omega-composites are
more prospective for obtaining negative real values of permittivity
and permeability than the split-ring resonators.

\section{Introduction}

V. G. Veselago studied in sixties \cite{Veselago} the electromagnetic
properties of materials whose permittivity and permeability
have negative real parts simultaneously.  He showed that these media exhibit
unusual effects like anomalous negative refraction of electromagnetic
waves and existence of backward waves (when the group velocity
and the Poynting vector are directed opposite with respect to the phase
velocity). This concept implies many other interesting features of
wave propagation that were reviewed in \cite{Veselago}.

These {\it Veselago} media do not exist in nature. The interest in
creating such media artificially, for optical applications, has been
indicated by J. Pendry in \cite{lens}. This kind of medium can be
obtained with arrays of conducting inclusions of a special shape. It
is obvious that the response of these materials will be resonant.
Therefore, negative refraction  is possible only within a more or
less narrow frequency band (or bands) \cite{freq}. This "negative"
band can be a part of a resonant band of inclusions that are used in
the composite. This is quite clear analyzing the Lorentz model of the
resonant dispersion.

At this stage of studies, since it is easier to use resonant
inclusions with millimeter dimensions, it seems to be reasonable to
focus on applications for the microwave region of the spectrum (optic composites
will need micrometer particles) and to study their response
experimentally. How to prepare these composites in a best way? Which
shape of inclusions is more suitable? These are the general questions
addressed by this article.

\begin{figure}
\centering \epsfig{file=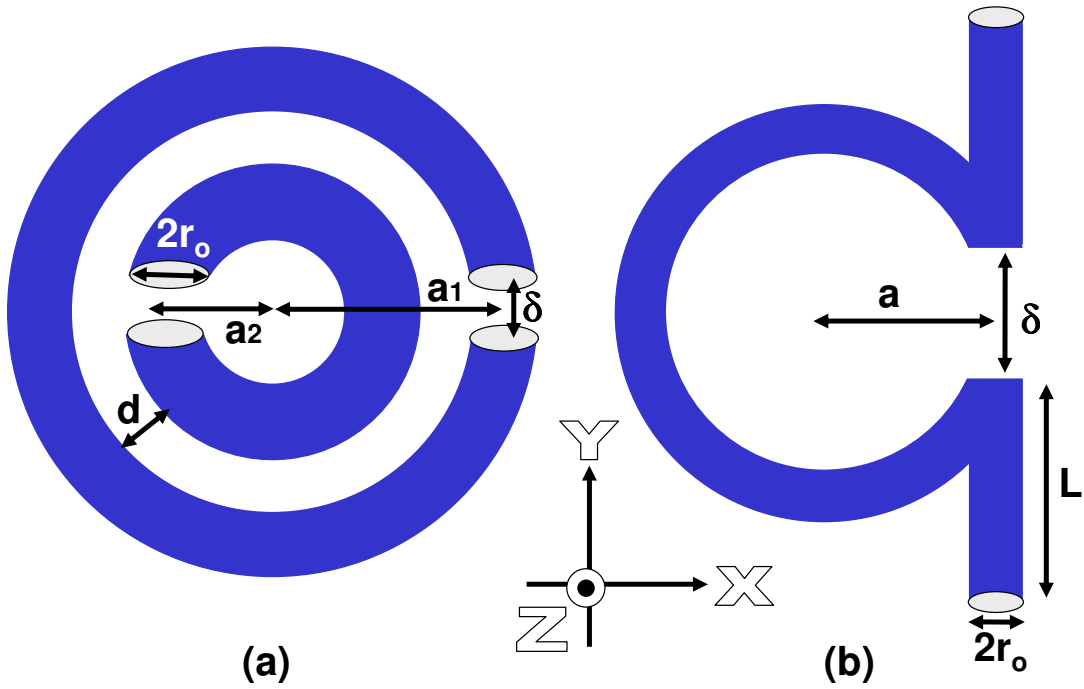, width=15cm} \caption{
Geometry of particles. (a) SRR particle  (b) Omega particle
 } \label{fig1}
\end{figure}

Works that have theoretically predicted negative permittivity for
composite media have been reviewed in \cite{scotland}. However,
publications containing the same results for the permeability are not
so common. In \cite{Pendry} and \cite{Smith} it was indicated that it
was a very difficult problem to obtain a magnetic resonant response
of inclusions (in order to get a negative real part of $\mu_{eff}$).
Probably, it was the reason for which these authors have chosen a rather
complicated particle named as {\it split-ring resonator (SRR)},
see Fig. \ref{fig1}. They maintained in these works, that SRR has
only a resonant magnetic polarizability and no resonant electric one.

In \cite{Smith} it was suggested that to obtain negative permittivity
simultaneously with  negative permeability.  SRR inclusions must be
combined with a lattice of straight wires. This complex system is a
very cumbersome technical solution. The only special case where wave
propagates normally to the wires (which corresponds to
\cite{Science}) is suitable to observe the negative refraction in
such a structure. It is clear that the structures from \cite{Smith}
cannot be considered as an {\it isotropic} backward-wave composite.

In this paper we show that the SRR particles alone allow to obtain simultaneously
negative permeability and permittivity. Indeed,
these inclusions also possess a resonant electric polarizability that
has not been discovered in \cite{Pendry} and \cite{Smith}. We also point
out that the analytical model of SRR suggested in \cite{Pendry} does
not give a satisfactory explanation of the frequency behavior of
effective permeability. Moreover, it is possible to realize an {\it
isotropic} composite with negative refraction choosing properly the
geometry, arrangement and concentration of SRR particles in a
host dielectric matrix.

However, we have found that SRR is not the best choice for obtaining a
backward-wave medium at microwaves. Actually, the resonant electric
response of SRR is weaker than the magnetic one. We show that it is better to
use an inclusion known as Omega particle (or $\Omega$-shaped
conducting particle) suggested in 1992 by Saadoum and Engheta
\cite{JE} since its magnetic and electric responses of  are of the
same order.

Numerical and analytical modeling of Omega particles have been
recently made in \cite{electromagn2}-\cite{other2} and the properties
of uniaxial and planar Omega composite media have been derived. It
appears that these media are prospective for creating low-reflecting
shields and antenna radomes in microwave applications. With this kind
of objectives, in 1995-1996, we studied theoretically uniaxial Omega
composites and noticed that sometimes the real parts of both
permittivity and permeability became negative, especially in the
cases with high densities of particles. Thus, as it was not suitable
for our goals, the experimental samples of Omega composites which
were prepared under guidance of S.A. Tretyakov \cite{electromagn2}
were dilute mixtures in order to avoid the apparition of negative
properties. Negative permittivity and permeability of Omega
composites has been predicted in the paper \cite{SE}. However, the
model of the material parameters presented in this work and based on
the transmission-line model of Omega particle was contradictory
(particle size was taken of the order $\lambda$ but the constitutive
relations were introduced as if the composites were continuous media)
and violated some physical foundations (Kramers-Kronig relations,
etc).

\begin{figure}
\centering \epsfig{file=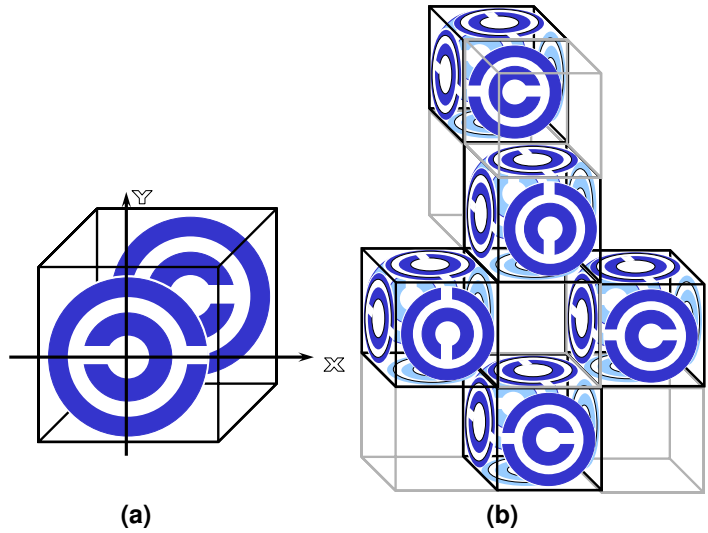, width=10cm}
\caption{Preparing the isotropic group of SRR particles. (a) A unit
cell contains 6 particles.  Particles located on opposite side of a
cubic cell are rotated by $180^o$. (b) Composite can be prepared
using the principle of chessboard. } \label{fig11}
\end{figure}

\begin{figure}
\centering \epsfig{file=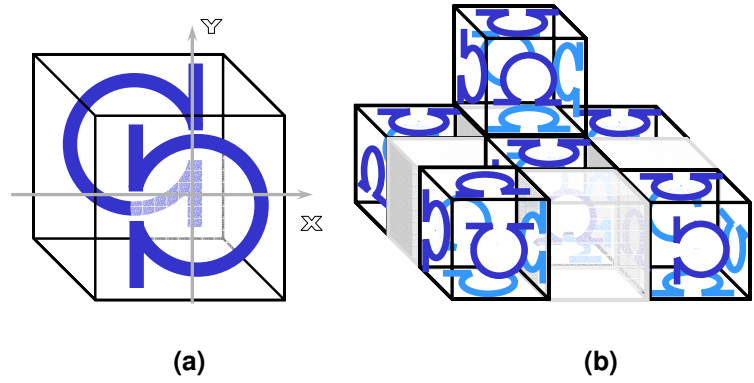, width=10cm}
\caption{Isotropic group of Omega particles (the same method as in
\ref{fig11}). (a) Omega particles are positioned at the sides of a
cubic unit cells. (b) The arrangement of unit cells shown here allows
to obtain an effectively isotropic medium.} \label{fig33}
\end{figure}

In the following we consider the effective properties of isotropic
composites of both Omega and SRR particles embedded in a dielectric
host medium. We study two different kinds of particles arrangement.
The first one is a mixture with a random orientations of particles. The
second one is obtained by positioning the particles at the sides of a
cubic unit cell, like it is shown in Figs \ref{fig11} and
\ref{fig33}. Different inclusions and structures are finally
compared.

\section{Theory}

Both SRR particles and Omega particles are bianisotropic particles. In
other words, an external electric field produces a magnetic dipole in
the particle and, at the same time, an external magnetic field
produces an electric dipole. For SRR, this magnetoelectric effect
is reduced by choosing the opposite positioning of the
splits of both loops. However, it cannot cancel out exactly
 since the loops are of different sizes.

Let a SRR particle be located in an infinite and homogeneous
dielectric medium with permittivity $\epsilon$. Consider the
electromagnetic response of an individual SRR particle to local
magnetic and electric fields.  The polarizabilities of a SRR particle
are dyadic coefficients relating the local electric and magnetic
fields $\-E^{\rm loc},\, \-H^{\rm loc}$ with the induced electric and
magnetic dipole moments:
\begin{eqnarray}
 \-p &=&\=a_{ee}\. \-E^{\rm loc}   +\=a_{em}\.  \-H^{\rm
loc}
\l{p} \\
\-m & =  & \=a_{me}\.  \-E^{\rm loc} +  \=a_{mm}\. \-H^{\rm loc}
\l{m}
\end{eqnarray}

The electric and magnetic moments of the SRR as a whole are vector
sum of the moments induced in the outer and inner loops (indices 1 and 2,
respectively) :
\begin{eqnarray}
 \-p & = &  \-p_1+ \-p_2\l{pp} \\
 \-m & = & \-m_1+ \-m_2\ \l{mm}
\end{eqnarray}
In this section we evaluate analytically the polarizabilities of a
SRR particle, assuming that the inclusion is fabricated from a round
wire of radius $r_0$. The section of the wire is small enough to
satisfy the inequality $r_0\ll a_{1,2}$, where $a_{1,2}$ are the
averaged radiuses of rings 1 and 2 (see Fig. \ref{fig1}a). Distance
$\delta$ between the broken ends of each ring is assumed to be small
compared to $a_{1,2}$. The analytical model of the Omega particle
will not be reproduced here. Analytical antenna theory and its
verification can be found in \cite{JEWA}.

Finally, using  the calculated polarizabilities we derive for each kind
of particles the material parameters through the well-known Maxwell
Garnett model.

\subsection{ SRR particle eigenfrequencies}

We consider the SRR particle as a pair of two mutually coupled
resonant scatterers. The complex amplitudes of induced currents at
the splits of both rings can be written as follows:
\begin{eqnarray}
I_1(Z_{C1}+Z_{L1})+ I_2Z_{12}&=&{\cal
E}_{H1}+{\cal E}_{E1}
,\l{i1} \\
I_2(Z_{C2}+Z_{L2})+ I_1Z_{12}&=&{\cal
E}_{H2}+{\cal E}_{E2} \l{i2}
\end{eqnarray}
Indices 1 and 2 refer to the outer and inner ring
respectively. ${\cal E}_{H}$ and ${\cal E}_{E}$ are the electromotive
forces induced by the external fields. $I_{1,2}$ are the currents at the
splits of the rings. $Z_{L1,L2}$ represent the inductive loop
impedances ($Z_{L1,L2}=j \omega L_{1,2}$), and $Z_{C1,C2}$ are the
capacitive impedance ($Z_{C1,C2}=1/ (j \omega C_{1,2})$) of the
slits. For a magnetic field excitation we can neglect the individual
loop capacitance which has only a small influence. For the study of
electric polarization,  influence of the proper capacitance of one loop
must be taken into account. For that we use the known theory of a split loop
antenna \cite{LoLee} and \cite{six} which takes into account the
capacitive properties of an individual loop scatterer and its
radiation losses. $Z_{12}$ in \r{i1},\r{i2} is  the mutual impedance
of broken loops. We have chosen to  include directly the capacitive
mutual coupling in the term $C_{1,2}$. Otherwise the analytical model
of mutual coupling would become too involved. With these
considerations the mutual impedance $Z_{12}$ in \r{i1},\r{i2} is
expressed only as a mutual inductance $M$ ($Z_{12}=j\o M$).

The capacitances $C_{1,2}$ are composed of two contributions.
The first part $C_S$ originates from the splits of each loop. The
second one is due to capacitive coupling of loops and can be seen
as a consequence of the following effect. Consider first as
a reference the conductor of one loop (the outer loop of a SRR
particle, for example). Next, place another conductor parallel to the
first one (in our case, the inner loop, with the split at the
opposite side). Then, assume that one half of the first loop (e.g.
the upper part) contains positive charge $+Q$, and the second half of
the first loop (e.g. the lower part) contains negative charge $-Q$.
These charges are distributed somehow along the loop perimeter. As
the two loops are positioned closely, in the upper half of the second
loop, a negative charge $-q$ is induced. Hence, the distribution of
the global charge $-q$ around the second loop repeats the
distribution of $+Q$ in the first loop. Note that we do not have
necessarily the correspondence $Q=q$. Of course, the same thing
happens in the lower part of the structure, where the induced charge
$+q$ appears. So two equivalent capacitances $C_{mut}$ arise from
this situation as shown in Fig. \ref{fig2}. Hence, we obtain an
effective additional capacitance which is equal to $C_{mut}/2$,
connected in parallel to the proper capacitance of the split. One can
note that this equivalent scheme can be improved taking into account
that there is an additional capacitance of the second slit.  However,
we are trying to derive a model of SRR as simple as possible and the
complexity, increased by this additional capacitance, is not
justified.

Values of inductances $L_{1,2}$, mutual inductance $M$ and total
capacitances $C_{1,2}$ are analytically calculated in Appendix.

To obtain the eigenfrequencies of the SRR particle, we let the
right-hand side of \r{i1},\r{i2} vanish and equate the determinant to
zero: \begin{equation} (Z_{C1}+Z_{L1})(Z_{C2}+Z_{L21})-Z_{12}^2=0
\l{eigen}\end{equation} Equation \r{eigen} can be written as
\begin{equation} \left({1\over j\o C_1}+j\o
L_1+r_1\right)\left({1\over j\o C_2}+j\o L_2+r_2\right)+\o^2 M^2=0
\l{eigen1}\end{equation} where $r_{1,2}$ are small resistances that
represent the radiation and the Joule losses of both loops. They lead
to a small imaginary correction term denoted as $\kappa(\o )$ in
equation \r{eigen2}: \begin{equation}
(\o^2-\o_-^2)(\o^2-\o_+^2)+j\kappa(\o )=0 \l{eigen2}\end{equation}
If we assume that the SRR particle is perfectly conducting, the term
$\kappa$ describes only the radiation losses of a magnetic dipole and
it can be derived from the energy balance condition, as it was done
in \cite{tretyakov} for arbitrary dipole scatterers. Comparing
\r{eigen1} with \r{eigen2} we find the eigenfrequencies:
\begin{equation} 2\left({\o_{\pm}\over \o_m}\right)^2={\o _m^2\over
\o _1^2}+{\o _m^2\over \o _2^2} \pm \sqrt{\left({\o _m^2\over \o
_1^2}+{\o _m^2\over \o _2^2}\right)^2-4} \l{pm}\end{equation} where
we denote
$$
\o _m^2={1\over \sqrt{C_1C_2(L_1L_2-M^2)} },\quad \o
_{1,2}^2={1\over C_{1,2}L_{1,2} }
$$

Frequencies \r{pm} correspond to two eigenmodes. If the
frequencies $\o _-$ and $\o _+$ differ dramatically the two eigenmodes of
the induced current are excited at different frequencies. This leads
to two separate resonances (that magnetic and that electric ones) of the
particle. However, we will see below that in practical cases $\o
_+\approx \o _-$. The two resonance bands overlap, and we obtain a
result close to that of the Lorentz model with only one resonance, as
it goes for simple scatterers. As a consequence, it matches
qualitatively the results of \cite{Pendry}.

\subsection{ Magnetic polarizability of SRR particle}

Next, we are going to derive the polarizabilities of SRR particle
using \r{i1},\r{i2} and definitions \r{p} and \r{m}. First, we study
the magnetic polarizability. For a magnetic excitation ($\-E^{\rm
loc}=0, \-H^{\rm loc}=\-z_0H^{\rm loc}$, where the $z$-axis is
perpendicular to the loop plane as it is shown in Fig. \ref{fig11})
we obtain: \begin{equation} {\cal E}_{H1,2}=-j\o \M0 S_{1,2} H^{\rm
loc}, \quad m_{1,2}=I_{1,2} \M0 S_{1,2} \l{def}\end{equation} Here
$S_{1,2}=\pi a_{1,2}^2$ are the effective loop areas.

The only component of the magnetic moment is vertical, i.e.
$$
\=a_{mm}=\-z_0\-z_0a_{mm}^{zz}
$$

Then, relations \r{def} together with \r{p}-\r{i2} and
\r{eigen}-\r{eigen2}  lead to the following expression:
\begin{equation} a_{mm}^{zz}=-D\o ^2
{B_1(\o^{2} -\o '^{2}_1)+B_2(\o^{2} -\o '^{2}_2)\over \Delta+j\kappa}
\l{amm} \end{equation} where we have denoted
\begin{equation} \Delta=(\o^2-\o_-^2)(\o^2-\o_+^2)\l{del} \end{equation} $$
B_{1}=S_1(S_1L_2-MS_2),\qquad B_{2}=S_2(S_2L_1-MS_1)
$$
$$
D={ \M0 ^2 \over L_1L_2 -M^2}
$$
and
 \begin{equation} \o '^2_{1,2}={\o^2 _{1,2}\over
\left(1-{MS_{2,1}\over L_{2,1}S_{1,2}} \right)}
\l{omprim}\end{equation} Now let us find the term $\kappa$. The
theory from \cite{tretyakov} based on the balance of received and
re-radiated power for any particle with magnetic dipole
polarizability $a_{mm}^{zz}$ gives the equation
\begin{equation} {\rm Im}{1\over a_{mm}^{zz}}={ k^3
\over 6\pi \M0} \l{kapp}\end{equation} Comparing this formula with
\r{amm} we come to the following expression for the term $\kappa$:
\begin{equation} \kappa={\o ^2 D k^3\left[B_1(\o^{2} -\o
'^{2}_1)+B_2(\o^{2} -\o '^{2}_2)\right] \over 6\pi \M0}
\l{kappa}\end{equation} With an appropriate calculation of
inductances and capacitances (see Appendix), it is now possible to
completely evaluate the magnetic polarizability $a_{mm}^{zz}$.

\subsection{ Electric polarizability of SRR particle}

Now we are interested in the electric polarizability.
Consider the action of local electric fields on a SRR. Two
possibilities exist to induce polarization on the SRR particle.

First, we study the case where the electric field is along the
horizontal direction ($x-$direction for an individual particle). The
only electro-dipole mode of the current is excited in both loops in
this case. The induced charges are concentrated at the splits and at
the opposite points and the polarizability component $a_{ee}^{xx}$ is
quasi-static. It does not depend on the presence or absence of the
split and equations \r{i1},\r{i2} are useless in this case\footnote{
Because $I_{1,2}=0$ and the electromotive force ${\cal E}_{E}$
induced at the split is also equal to zero: charges are the same at
the edges of each split.}.

In \cite{IEEE} we derived formula (28) for $xx-$component of the
polarizability of a circular loop with area $S=\pi a^2$. Denoting
this polarizability as $a_{QS}$ we obtain from \cite{IEEE}:
$$
a_{QS}=-{4SJ'_1(ka)\over \o \eta
A_1(ka)}
$$
In this equation,  $\eta=\sqrt{\M0/\epsilon\E0}$ is the wave
impedance of the host medium, $k=\o \sqrt{\M0\epsilon\E0}$ is the
wave number in the medium, $J'_1(ka)$ is the derivative of the
Bessel function of the first order, $A_1(ka)$ is one of so-called King's
coefficient known from the theory of loop antennas
\cite{six},\cite{LoLee}. King's coefficients with a very high accuracy
are approximated in \cite{six} and take the following form:
$$
A_1(ka)= \left({ka-{1\over ka}\over \pi}\right)
\left(\log{8a\over r_0}-2\right)
-{0.667(ka)^3-0.207(ka)^5\over \pi}$$
$$
-j0.333(ka)^2-j(0.133ka)^4+j0.026(ka)^6
$$
$$
A_0(ka)={ka\over \pi}\left(\log{8a\over r_0}-2\right)
 +{0.667(ka)^3-0.267(ka)^5\over \pi}
-j0.167(ka)^4-j0.033(ka)^6
$$
$$
A_2(ka)=\left(ka-{4\over ka}\over \pi\right)
\left(\log{8a\over r_0}-0.667\right)
+{-0.4ka+0.21(ka)^3-0.086(ka)^5\over \pi}
$$
$$
-j0.05(ka)^4-j0.012(ka)^6
$$

Since we have two loops, we must sum up their $a_{QS}$ to obtain the
result: \begin{equation} a_{ee}^{xx}=-{4S_1J'_1(ka_1)\over \o \eta
A_1(ka_1)}-{4S_2J'_1(ka_2)\over \o \eta A_1(ka_2)}
\l{aex}\end{equation} Now we consider the case where the electric
field is directed along $y-$axis. Here the resonant electric
polarization appears in both loops of the SRR particle. To obtain the
resulting polarizability, we combine the theory of a single loop with
resonant electric polarization (see \cite{LoLee} and \cite{IEEE})
with equations \r{i1} and \r{i2}.

In \cite{IEEE} the following formula has been derived for the
$yy-$component of the electric polarizability of a single loaded loop
of radius $a$: \begin{equation} a_{ee}^{yy}= {4\pi a^2J'_1(ka)\over
\o \eta A_1(ka)}\left(1 + {2j\over \pi \eta A_1(ka)}   \    {1\over
Y_{L}+Y_{\rm split}} \right) \l{xixi}\end{equation} Here $Y_{\rm
split}$ is the admittance that is connected to the loop split. For
both loops we have $Y_{\rm split} \equiv 1/Z_{\rm split}= j\o
C_{1,2}$ which include the mutual capacitance of the loops.
$Y_L=1/Z_L$ is the proper loop admittance. In our case of coupled
loops, we should replace $Y_L$ by the value $Y_{1,2}$ which takes
into account their mutual inductance. For loop $1$, we can say that
it is composed of the proper impedance of the loop $Z_{L1}$ and of
the additional impedance $Z_{12}I_2/I_1$, which is the contribution
of loop $2$ into loop $1$. Then, for loop $1$, we have to substitute
into \r{xixi}:

$$
{1\over Y_1}=Z_{1}=Z_{L1}+Z_{12}{I_2\over I_1}
$$
Similarly, for the second loop we have:
$$
{1\over Y_2}=Z_{2}=Z_{L2}+Z_{12}{I_1\over I_2}
$$

Denoting the ratio $I_1/I_2$ as $\xi$ and taking into account that
$Z_{12}=j\o M$, we come to the following formulas for the
$yy-$polarizabilities of both loops: \begin{equation} a_{e1}^{yy}=
{4S_1J'_1(ka_1)\over \o \eta A_1(ka_1)}\left(1 + {2j\over \pi \eta
A_1(ka_1)}{1\over Y_{1}+j\o C_1} \right) \l{xi1}\end{equation}
\begin{equation} a_{e2}^{yy}= {4S_2J'_1(ka_2)\over \o \eta
A_1(ka_2)}\left(1 + {2j\over \pi \eta A_1(ka_2)}{1\over Y_{2}+j\o
C_2} \right) \l{xi2}\end{equation} where
$$
Y_{1}={Y_{L1}\over
1+{j\o MY_{L1}\over \xi}
}
$$
$$
Y_{2}={Y_{L2}\over
1+j\o MY_{L2} \xi
}
$$
The proper loop admittance is given by \cite{six}: \begin{equation}
Y_{L}={1\over j\pi\eta}\left( {1\over A_0(ka)}+ {2\over
A_1(ka)}+{2\over A_2(ka)} \right) \l{proper}\end{equation}
Expressions of $Y_{L1,L2}$ follow from \r{proper} with substitutions
$a=a_{1,2}$.

Equations \r{i1},\r{i2} allow to find $\xi$:
\begin{equation} \xi={{\cal E}_{E1}(Z_{L2}+Z_{C2})-{\cal E}_{E2}Z_{12}
\over {\cal E}_{E2}(Z_{L1}+Z_{C1})-{\cal E}_{E1}Z_{12} }
\l{xifi}\end{equation} If a loop with proper impedance $Z_L$ loaded
with the impedance of the split $Z_{\rm split}$ is submitted to an
external electric field, an electromotive force is induced. Its value
is given by formula (26) from \cite{IEEE}: \begin{equation} {\cal
E}_{E}= {4aJ'_1(ka)({1\over Y_L}+{1\over Y_{\rm split}} )\over j \eta
A_1(ka)}\left( 1+ {j\over \pi \eta}  \  {1\over Y_L+Y_{\rm split}}
\right)E_y^{\rm loc} \l{eee}\end{equation} Substituting the values
$a_{1,2}$ and $r_0$ into \r{proper} and \r{eee} allows to calculate
the electromotive forces induced by an external electric field in
each loop of the SRR particle. Then, substituting expressions of
${\cal E}_{E1,E2}$ into \r{xifi} with the help of relation \r{proper}
we can find $\xi$. Consequently, we derive an expression of $Y_{1,2}$
which is then substituted into \r{xi1} and \r{xi2}. Finally, we
evaluate $a_{ee}^{yy}$ as the sum
\begin{equation} a_{ee}^{yy}=a_{e1}^{yy}+a_{e2}^{yy} \l{reson}\end{equation}

At this point, we have at our disposal a complete analytical model of
the electric and magnetic polarizabilities of an individual SRR
particle (other components of $\=a_{ee}$ and $\=a_{mm}$ are zeros).

\subsection{Effective polarizabilities of a planar bianisotropic particle
in isotropic mixtures}

An isotropic mixture with a random orientation of particles can be
considered as a mixture realised with  isotropic particles which have
averaged scalar polarizabilities \cite{MG}: \begin{equation}
a_{ee}^{av}=\sum\limits_{\alpha=1}^3 {a_{ee}^{\alpha\alpha} \over
3},\qquad a_{mm}^{av}=\sum\limits_{\alpha=1}^3 {a_{mm}^{\alpha\alpha}
\over 3},\quad a_{em}^{av}=a_{me}^{av}=0 \l{scalar}\end{equation}
Here we have designated Cartesian components with index
$\alpha=x,y,z$. The particles under consideration do not have cross
components: $a_{ee}^{\alpha\beta}=0$ for $\alpha\ne \beta$. Moreover,
we have seen with our analytical model, that the only
polarizabilities that are non-zero are: $a_{ee}^{xx},a_{ee}^{yy}$ and
$a_{mm}^{zz}$.

If SRR or Omega particles are arranged on the sides of cubic unit
cells (see Figs. \ref{fig11} and \ref{fig33}), the magnetoelectric
effect also disappears and dyadics $\=a_{em},\=a_{me}$ of the
individual particles do not contribute into material parameters.
Furthermore, mutual coupling of six particles belonging to the same
cell is weak enough if the cell size $d_{cell}$ is a bit more than
the particle size $d_p$ (the particles do not touch one another).
Now, let us evaluate the equivalent effective polarizability of each
kind of unit cell ignoring the mutual coupling between the particles
of a cubic cell.

\subsubsection{Effective polarizabilities of a cubic cell of SRR particles}

The cubic cell under consideration is presented Fig. \ref{fig11}. An
electric field directed along an arbitrary edge of a cubic cell
excites the electric dipoles in four SRR particles. For two particles
$\-E^{\rm loc}$ is parallel to the line connecting the rings splits
and corresponds to the polarizability $a_{ee}^{xx}$. For the other two
particles $\-E^{\rm loc}$ is perpendicular to this line and
corresponds to $a_{ee}^{yy}$.

Consider now the magnetic field effect on the cell. A magnetic field
directed along an arbitrary edge of a cell induces two equivalent
magnetic moments in two opposite particles. So, the magnetic
polarizability of a unit cell turns out to be the double of
$a_{mm}^{zz}$.

To summarize, the cubic unit cell is equivalent to an isotropic
particle with electric and magnetic polarizabilities given by
\begin{equation} a'_{ee}=2a_{ee}^{xx}+2a_{ee}^{yy}, \qquad
a'_{mm}=2a_{mm}^{zz} \l{splits}\end{equation}
\subsubsection{Effective polarizabilities of a cubic cell of Omega particles}

For a unit cubic cell of Omega particles (see Fig. \ref{fig33}) the
situation is quite different. Here the resonant excitation by the
electric field directed along the arms of $\Omega$ (along $y$-axis in
the model of an individual particle) is mainly due to the presence of
the arms \cite{JEWA}. Therefore the total electric polarization of
two opposite omega particles keeps resonant behavior. Within the
resonant band the quasi-static component of Omega particle
polarizability $a_{ee}^{xx\Omega}$ describing the response of a
particle to the electric field applied normally to the arms (see Fig.
\ref{fig1}) is small compared to $a_{ee}^{yy\Omega}$. Then one
obtains for a unit cell from Omega particles approximate
relations

\begin{equation} a'_{ee}\approx 2a_{ee}^{yy\Omega}, \qquad
a'_{mm}=2a_{mm}^{zz\Omega} \l{omegas}\end{equation} where
$a_{ee}^{yy\Omega},a_{mm}^{zz\Omega}$ can be taken from \cite{JEWA}.

\subsubsection{Material parameters of composites}

The material parameters of the isotropic mixture are given by the
standard Maxwell Garnett formulas \cite{MG}: \begin{equation}
\epsilon _{eff}=\epsilon + F \left( {Na'_{ee}\over
\E0}-{N^2a'_{ee}a'_{mm}\over 3\E0\M0} \right) \l{eps}\end{equation}
\begin{equation} \mu _{eff}=1 + F \left( {Na'_{mm}\over
\M0}-{N^2a'_{ee}a'_{mm}\over 3\E0\M0\epsilon} \right)
\l{mu}\end{equation} In these equations
$$
F^{-1}={1-{Na'_{ee}\over{3\E0\epsilon}}}-{Na'_{mm}\over{3\M0}}+
{N^2a'_{ee}a'_{mm}\over{9\E0\M0\epsilon}}
$$

These equations are general for the different kinds of composites we
are studying. Depending on the composition of mixture, only expressions for
polarizabilities change.

\begin{itemize}
\item
for {\it random media}, which are realized by mixing particles in an
isotropic and random way in a host medium, $N$ denotes the
concentration of particles, and $a'_{ee}=a_{ee}^{av}$,
$a'_{mm}=a_{mm}^{av}$ are given by \r{scalar}
\item
for {\it cubic-cell media}, which are obtained with previously mentioned
cubic unit cells,  $N$ denotes the concentration of cells, and
$a'_{ee}$ and $a'_{mm}$ are given by \r{splits} or \r{omegas}.
\end{itemize}

\section{Model validation for SRR particle}

Our analytical model is below compared with numerical simulations. Numerical
model which allows to simulate the averaged polarizabilities of an
arbitrary bianisotropic particle with thin wire of round cross
section, that allows to calculate  also  material parameters of a
random composite has been derived from \cite{Bruno2}, \cite{Bruno}.
Since its publication this method has been widely used to simulate
bianisotropic composites.

We compared the analytical and numerical models of the SRR prepared
with wires of radius $r_0=0.2$ mm. Numerical example presented here
corresponds to the following parameters. The averaged radiuses of the
inner and outer loops are equal to $a_2=2.1$ and $a_1=2.7$ mm. The
length of the broken part is taken $\delta=0.1$ mm or $\delta=0.4$ mm
($\delta$ is the same in both loops). The host medium permittivity
$\epsilon=1.5$.

In Fig. \ref{reamm} the real part of $a_{mm}^{av}$ is shown
calculated in both analytical and numerical ways for the case
$\delta=0.1$ mm. The imaginary parts of $a_{mm}^{av}$ for this case
are shown in Fig. \ref{imamm}. The real and imaginary parts of
$a_{ee}^{av}$ calculated in both ways are shown in Figs. \ref{reaee}
and \ref{imaee}, respectively.

Comparing the resonant frequencies of $a_{ee}^{av}$ and  $a_{mm}^{av}$,
we can note that the difference is quite negligible. This means that
the resonant bands overlap ($\o _+\approx \o _-$) as we mentioned
previously.

We can observe also, that the analytical model shows a weak
additional resonance for the electric polarization around 5.2 GHz.
This is a defect of our analytical  theory which is related to the
adopted approximations. We think that the second resonance follows
from our model of loops mutual coupling which is not accurate enough.
Adding a very small imaginary value to the right-hand side of formula
\r{M1} we completely suppress  this second  resonance without
changing the values of $a_{mm}$ and $a_{ee}$ within the first
resonant band. Of course, this correction has nothing to do with the
theory but it confirms that the model of $Z_{1,2}$ is not accurate
enough and can be improved.

The electric polarizability calculated in both analytical and
numerical ways for the case
$\delta=0.4$ mm (the other parameters are the same as above)
is presented in Fig. \ref{Brunore}.

The agreement between two models is quite the same for $\delta=0.1$
and $\delta=0.4$ mm. The same remarks could be made for $a_{mm}$.

Polarizabilities of SRR particle are very sensitive to deviations
of the particle parameters, and we can consider that the agreement
between the two models is satisfactory. It makes our analytical
theory suitable to calculate the effective composite parameters and,
hence, allows to solve explicitly the inverse problem for such
composites.

\section{SRR and Omega isotropic composites}

In this section we analyze the material parameters of effective media
obtained with SRR particles or Omega particles. The mixture is
fabricated by
\begin{itemize}
\item
distributing the particles with random orientations in a host medium
\item
positioning the particles at the sides of a cubic unit cell (filled by
the same medium as the matrix) as shown in Fig. \ref{fig11} and
\ref{fig33}.
\end{itemize}

\subsection{SRR composites}

In Figs. \ref{reeps} and \ref{remu} both $\epsilon_{eff}$ and
$\mu_{eff}$ of a {\it random SRR composite} are shown (calculated
in both analytical and numerical ways). Particles geometry and
background permittivity has been taken the same as above (the split width
is $\delta=0.4$ mm). The concentration of particles has been taken
$N=1$ particle/cm$^3$. For the particle maximal size $5.8$ mm this
concentration allows to use the Maxwell Garnett averaging procedure.

We can observe on Fig. \ref{reeps} that the permittivity is not
becoming negative within the resonant band (whereas both models of
the permeability allow the negative values for ${\rm
Re}(\mu_{eff})$). Here, the permittivity of SRR composite remains
positive because of the restriction of dilute mixtures (imposed by
the validity of the Maxwell Garnett model).

In the case of {\it SRR cubic media}, since we apply the Maxwell
Garnett model to the unit cells considering them as particles, we can
take the concentration of SRR particles greater than in the case of a
random mixture, and that, without violating the restriction of dilute
mixtures.

As a consequence, in Fig. \ref{cub2} we can see that ${\rm
Re}(\mu_{eff})$ and ${\rm Re}(\epsilon_{eff})$ are negative
simultaneously. The concentration of cells has been calculated for
the case of unit cells arrangement as presented in Fig. \ref{fig33}.
The size of the cell is equal to $6\times 6\times 6 $ mm$^3$, and the
averaged distance between the centers of two adjacent cells is close to
$11$ mm. This allows to consider the composite with such cells as a
dilute mixture.

\subsection{Omega composites}

In Fig. \ref{fig43} and \ref{fig44} we present the results calculated
within the frame of the Maxwell Garnett model for an Omega composite
prepared from cubic units as shown in Fig. \ref{fig33}. In that case,
the maximal size of the Omega particle is equal to $6.4$ mm. The cell
size is equal to $6.5\times 6.5\times 6.5$ mm$^3$. For an illustration,
we have chosen the following dimensions. The radius of the loop is
$3$ mm and the radius of the wire is $r_0=0.1$ mm. The length of an arm
is $L=3$ mm. The host medium permittivity is $\epsilon=2$. Note that
for cases when $\delta \ll a$,  parameter $\delta$ does not influence
the polarizabilities at all (see \cite{JEWA}).

In Fig. \ref{fig43}, the real parts of the permittivity and permeability
of the Omega composite versus the frequency are shown. In Fig.
\ref{fig44}, the imaginary parts of permittivity and permeability are
presented.
We can see that ${\rm Re}(\epsilon_{eff})$ and ${\rm Re}(\mu_{eff})$ reach $-1$
practically at the same frequency. This is a very exciting result if
we refer to  paper \cite{lens} where the perfect lens has been
predicted for the case $Re(\epsilon_{eff})=Re(\mu_{eff})=-1$.

The case of {\it random Omega composite} (Fig. \ref{random_omega})
with density $N=1$ particle/cm$^3$ only differs by the heights of the
resonant peaks of the permittivity.

\section{Conclusion}

Based on the analysis of electromagnetic phenomena that appear in a
{\it split ring resonator}, we have developed an analytical model of
SRR particles. We have shown that the SRR particles can exhibit, on
their own, simultaneously  magnetic and electric polarizability. The
proposed model has been successfully compared with a numerical
approach. So, these results point out that previous works on SRR do
not give a satisfactory explanation of the behavior of this
particle. Next, we have calculated with the Maxwell Garnett approach the
effective parameters of composites realized with SRR particles. We
have shown that it is possible to obtain an isotropic composite with
negative refraction choosing properly the geometry, arrangement
and concentration of SRR particles in a host dielectric matrix.
The way of making these composites has been presented. With this
technique, we have obtained an effective material with simultaneously
negative real part of permittivity and permeability.

However, we have shown that SRR is not the only and  not the best
choice for obtaining a backward-wave medium at microwaves. Actually,
we can obtain the same kind of medium, using Omega particle instead
of SRR particles. With these inclusions we have derived a material
which gives ${\rm Re}(\epsilon_{eff})={\rm Re}(\mu_{eff})=-1$ in the
resonant band. This is the wished result exposed by the inventor of
the SRR particle. Finally, we can conclude that the Omega particle is
a more appropriate choice in order to create microwave composites
with negative refraction.

We have to point out yet, that our model exhibits a high level of
radiation losses within the resonant band. As a consequence, if
material samples are so lossy, it would be problematic to measure the
negative refraction on it. These high radiation losses are due to the
Maxwell Garnett model which implies a random organization of
particles and is consistent with the Kramers Kronig relations.
However, the proposed {\it cubic composite} can be classified in a
middle position between random media and lattices.  The composite is
constructed as a lattice, stacking up cubes in a tree dimensional
chessboard assembly, but it is not a perfect lattice. There is
indeed, a certain random organization in the positions of the
particles on the sides of the cubic cells (as it can be seen on
figures \ref{fig11} and \ref{fig33}). So, we can reasonably expect
that losses given by the Maxwell Garnett model will be overestimated.

\section{Appendix}

Inductance of a round loop of radius $R$ made with a wire of
cross section radius $r_0$ can be found e.g. in \cite{LoLee}:
 \begin{equation} L= \M0
R\left(\log\left({8R\over r_0}\right)- 2\right) \l{LL}\end{equation}
Inductances of both rings of SRR can be obtained with \r{LL} with the
substitution $R=a_{1,2}$.

Formula (5.23) from \cite{Kal} for two concentric loops whose radii
$a_1, a_2$ are close to one another, gives the following
approximation for the mutual inductance: \begin{equation} M= \mu_0
a_1 \left[(1-\chi)\log\left({4\over \xi}\right)-2+\xi \right]
 \l{M1}\end{equation} where $\chi=(a_1-a_2)/2a_1$ is a small
parameter ($0<\chi<0.25$). In \r{M1} we have neglected the terms of
order $\xi^2$.

\begin{figure}
\centering \epsfig{file=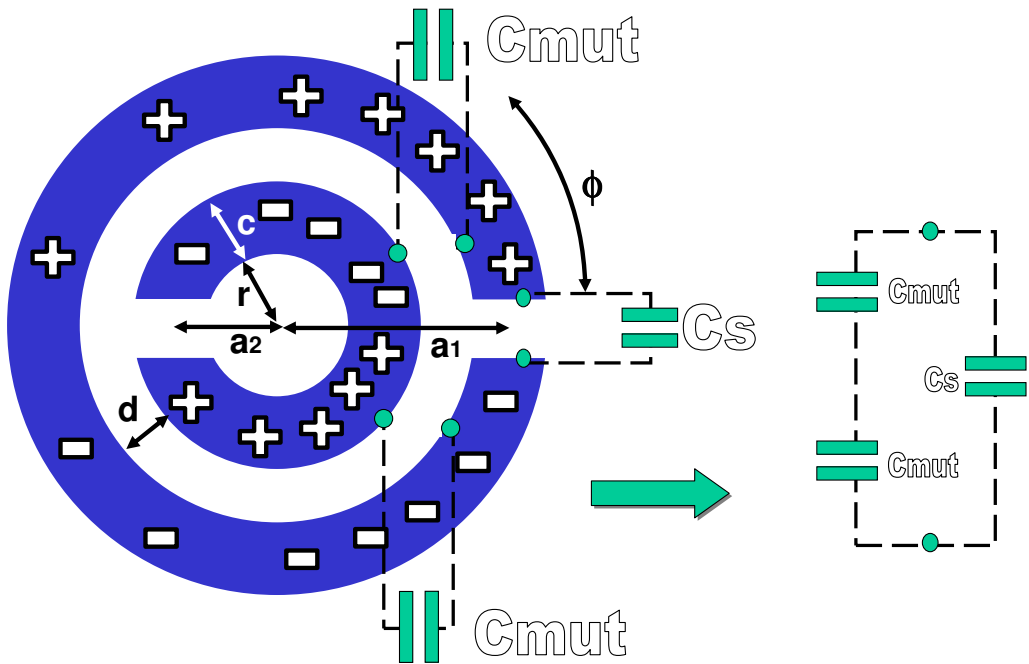, width=15cm}
\caption{Capacitive mutual influence of loops. The connection of
equivalent capacitors $C_{mut}$ is shown.} \label{fig2}
\end{figure}

\begin{figure}
\centering \epsfig{file=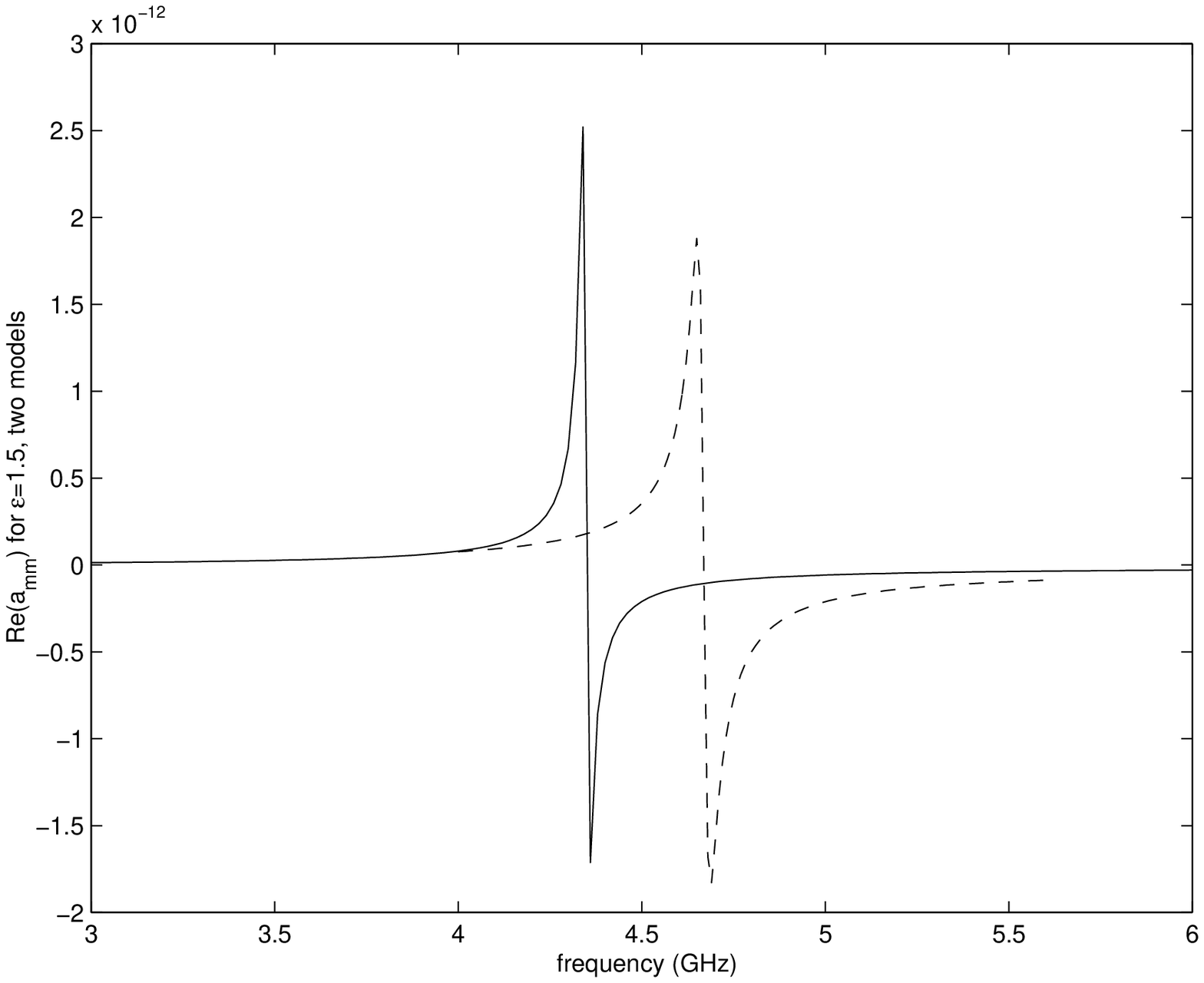, width=10cm} \caption{Real
part of the averaged magnetic polarizability of SRR versus frequency.
The split width $\delta$ is $0.1$ mm. Analytical model (solid) and
numerical model (dashed).} \label{reamm}
\end{figure}

\begin{figure}
\centering \epsfig{file=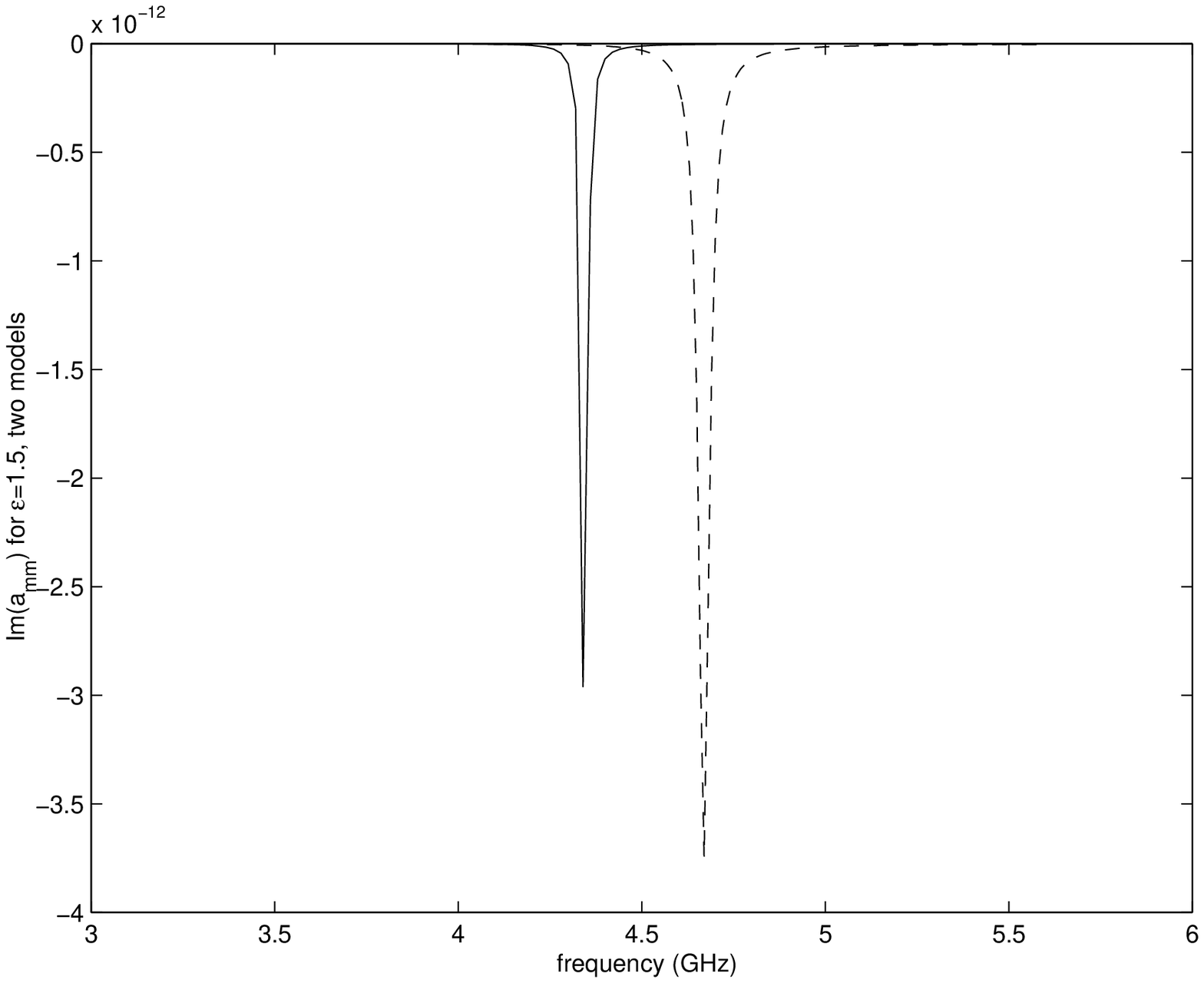, width=10cm}
\caption{Imaginary part of the averaged magnetic polarizability of
SRR versus frequency. The split width $\delta$ is $0.1$ mm.
Analytical model (solid) and numerical one (dashed).} \label{imamm}
\end{figure}

\begin{figure}
\centering \epsfig{file=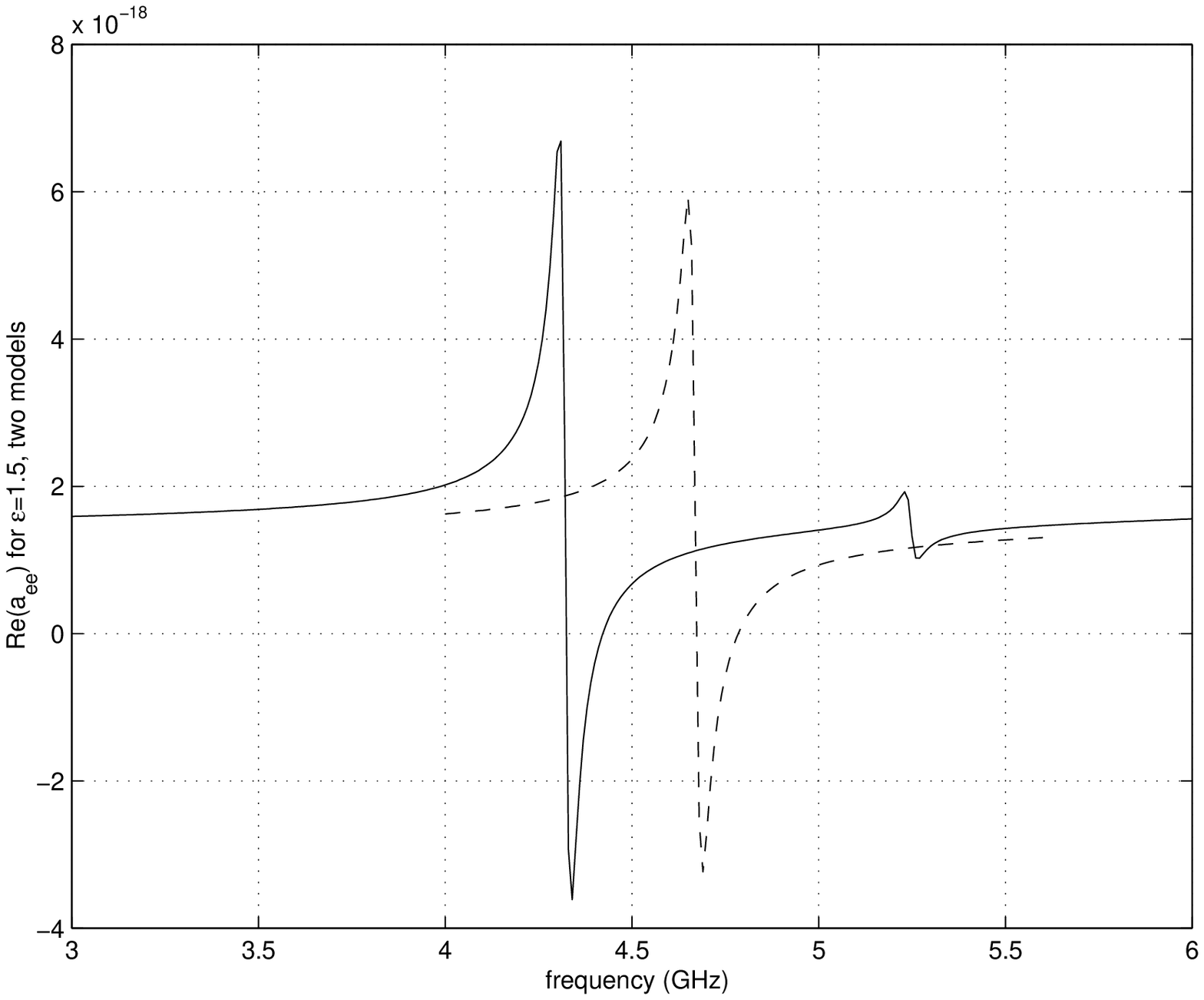, width=10cm} \caption{Real
part of the averaged electric polarizability of SRR versus frequency.
The split width is $0.1$ mm. Analytical model (solid) and numerical
model (dashed).} \label{reaee}
\end{figure}

\begin{figure}
\centering \epsfig{file=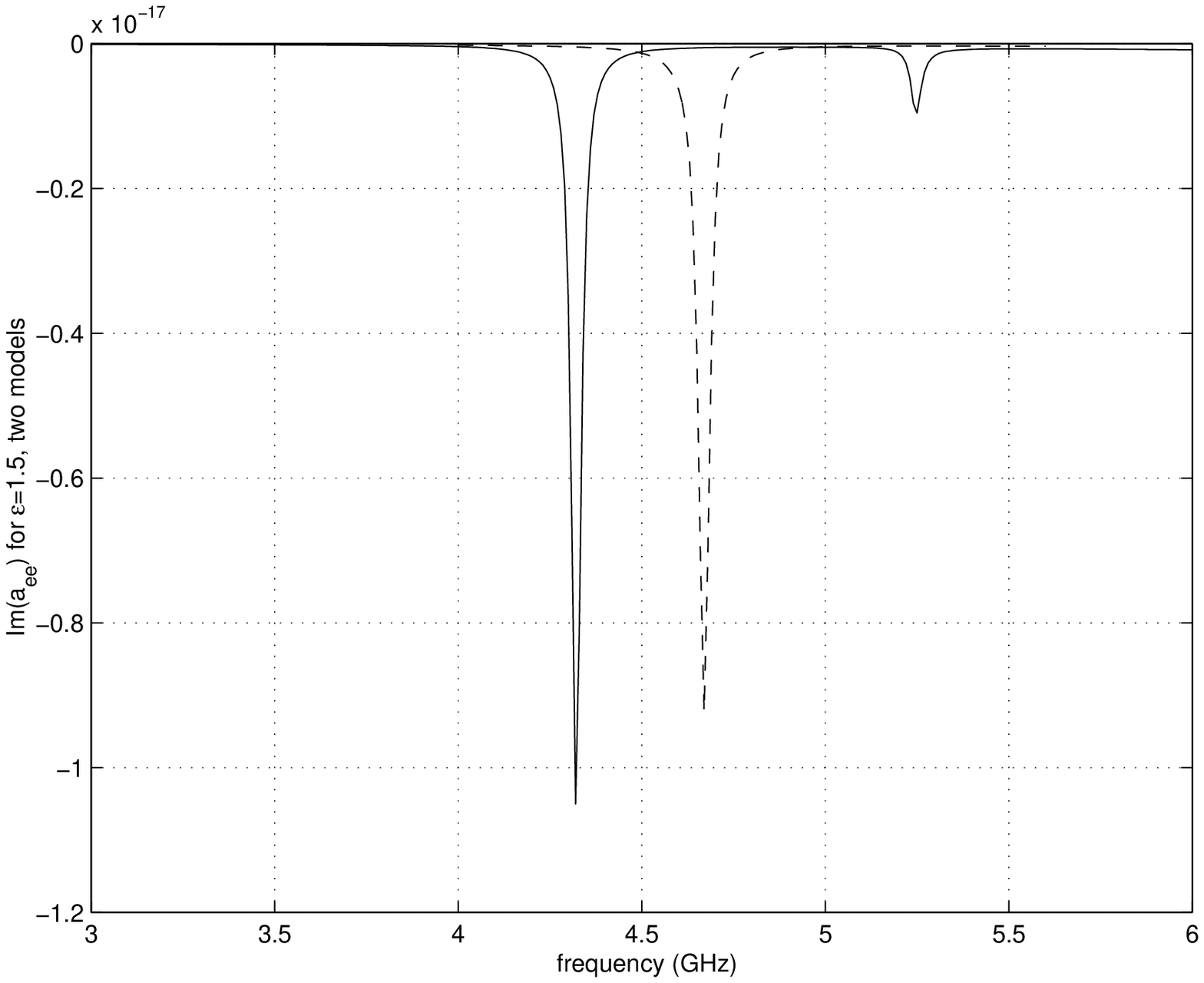, width=10cm}
\caption{Imaginary part of the averaged electric polarizability of
SRR versus frequency. The split width is $0.1$ mm. Analytical model
(solid) and numerical one (dashed).} \label{imaee}
\end{figure}

\begin{figure}
\centering \epsfig{file=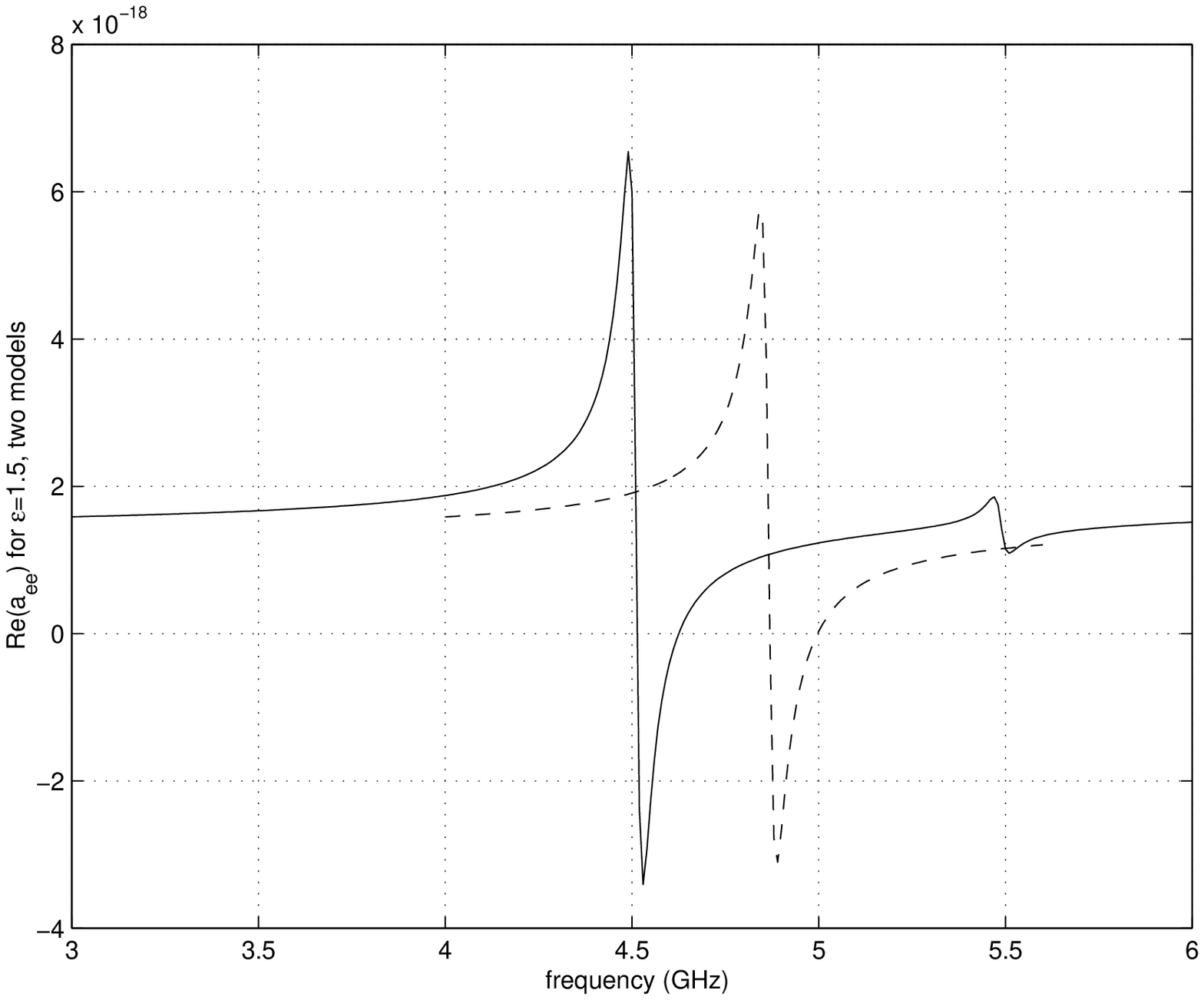, width=10cm} \caption{Real
part of the averaged electric polarizability of SRR versus frequency.
The split width is $0.4$ mm. Analytical model (solid) and numerical
one (dashed).} \label{Brunore}
\end{figure}

\begin{figure}
\centering \epsfig{file=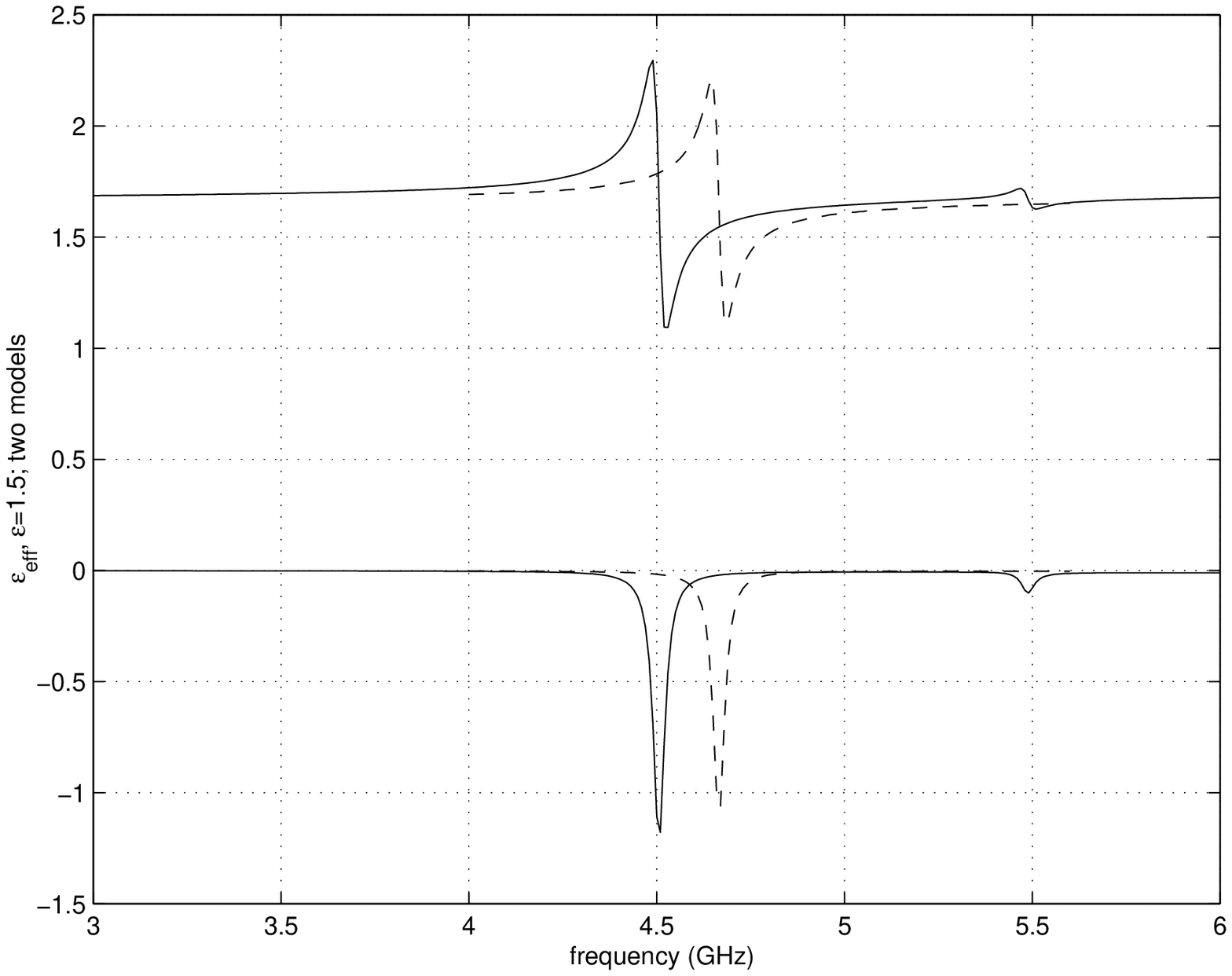, width=10cm} \caption{Real
and imaginary parts of the permittivity of the random SRR composite
versus frequency. Analytical model (solid) and numerical model
(dashed).} \label{reeps}
\end{figure}

\begin{figure}
\centering \epsfig{file=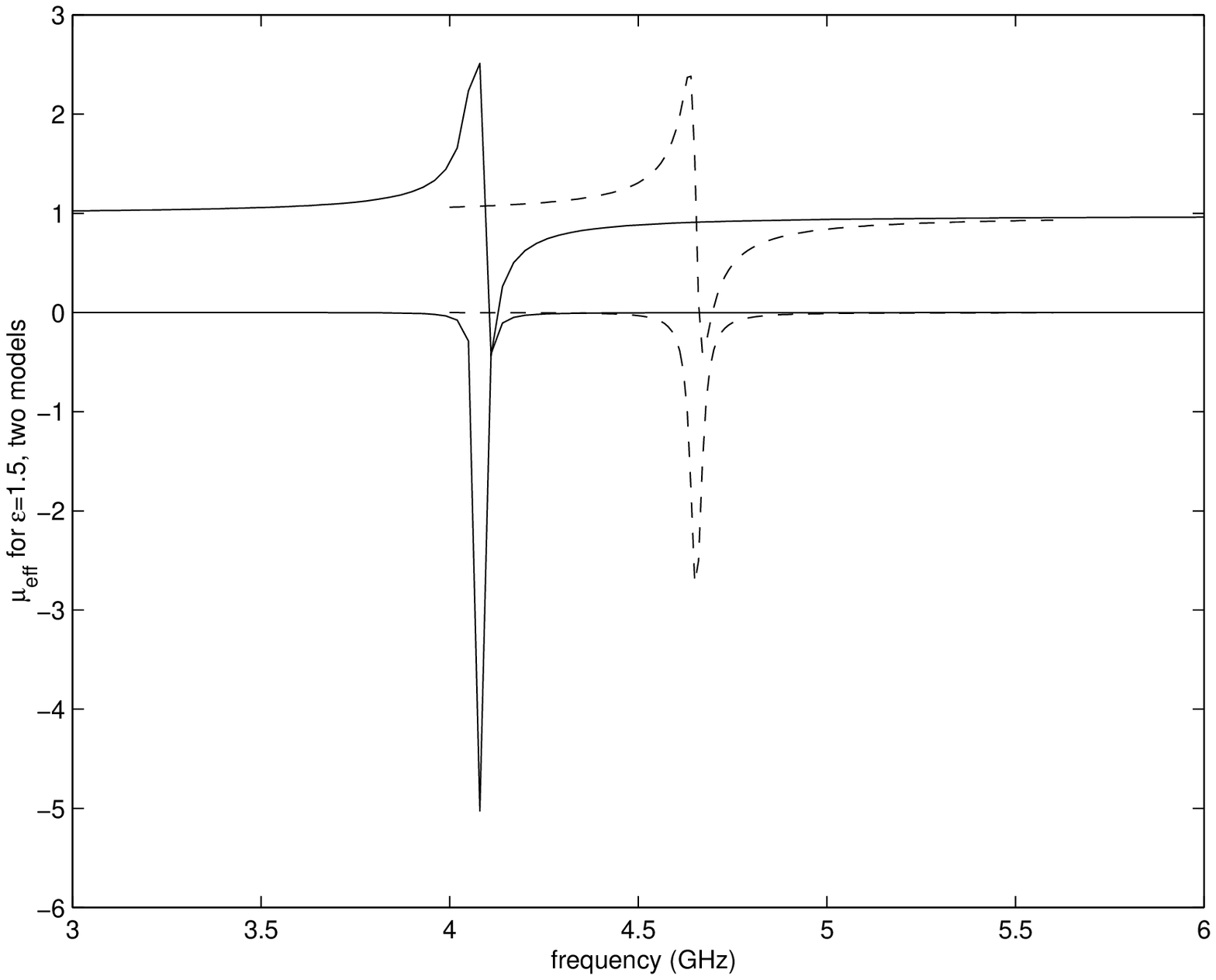, width=10cm} \caption{Real
and imaginary parts of the permeability of the random SRR composite
versus frequency. Analytical model (solid) and numerical one
(dashed).} \label{remu}
\end{figure}

\begin{figure}
\centering \epsfig{file=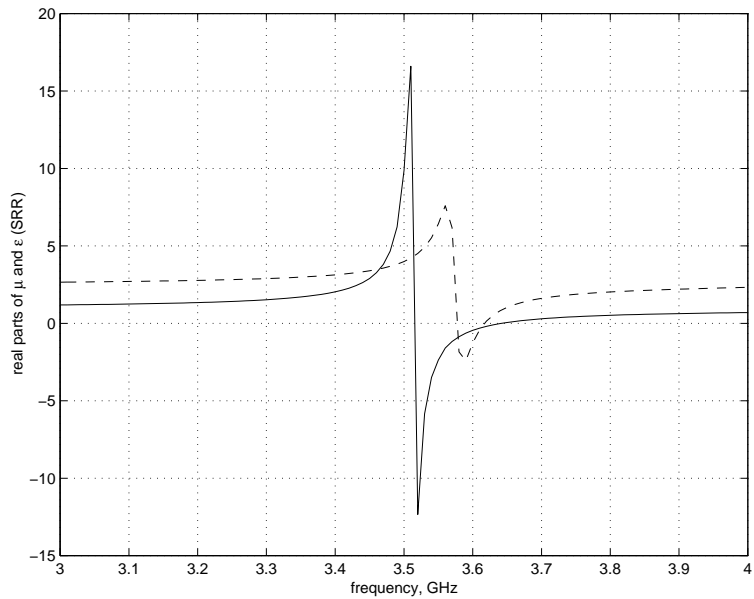, width=10cm} \caption{Real
parts of the permeability (solid) and permittivity (dashed) of the
SRR composite from cubic cells versus frequency. Analytical model.}
\label{cub2}
\end{figure}

\begin{figure}
\centering \epsfig{file=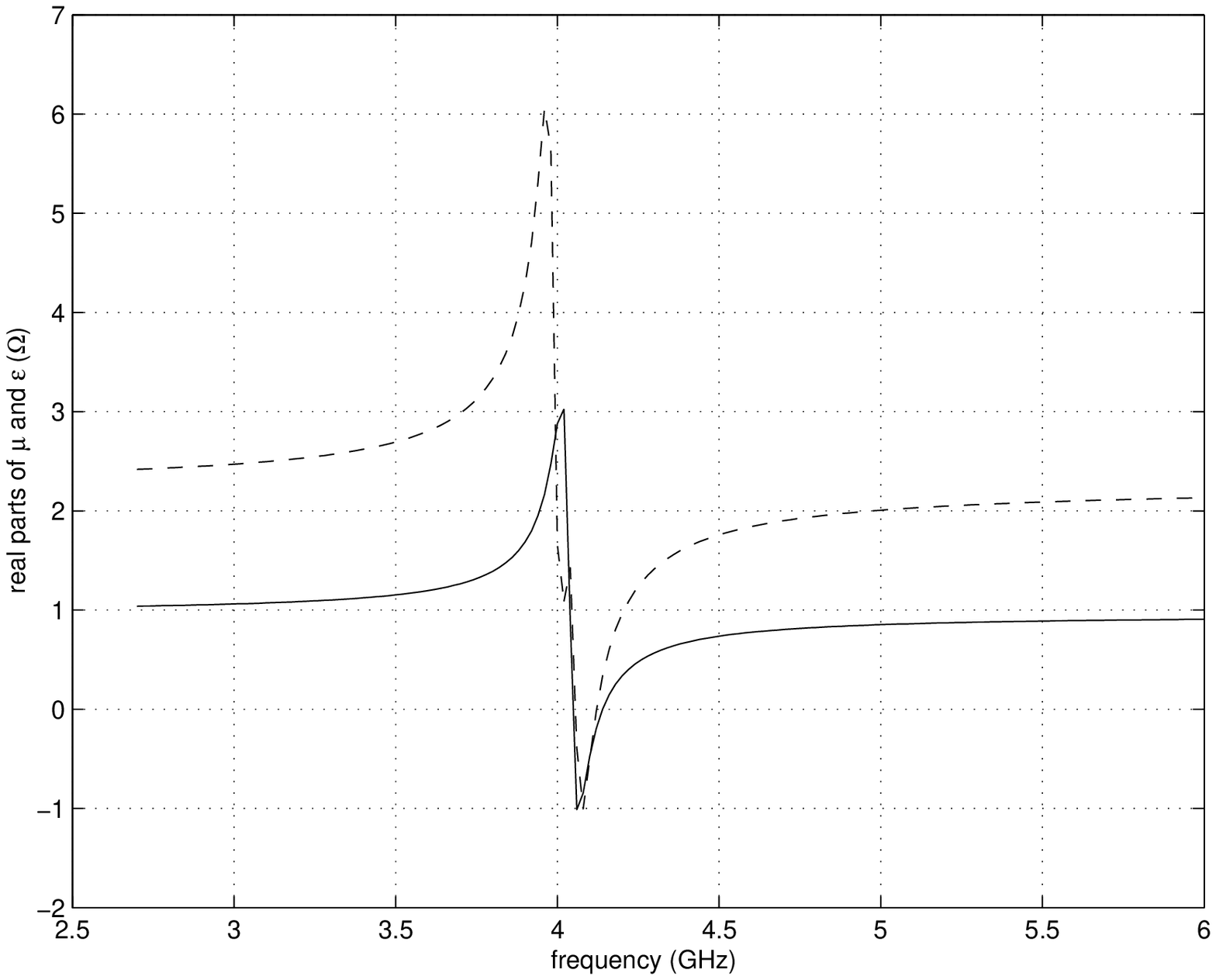, width=10cm} \caption{Real
parts of the effective permeability (solid) and permittivity (dashed)
of an Omega composite formed by cubic cells versus frequency. The
host permittivity is $2$  and the length of a unit cell is $6.5$ mm}
\label{fig43}
\end{figure}

\begin{figure}
\centering \epsfig{file=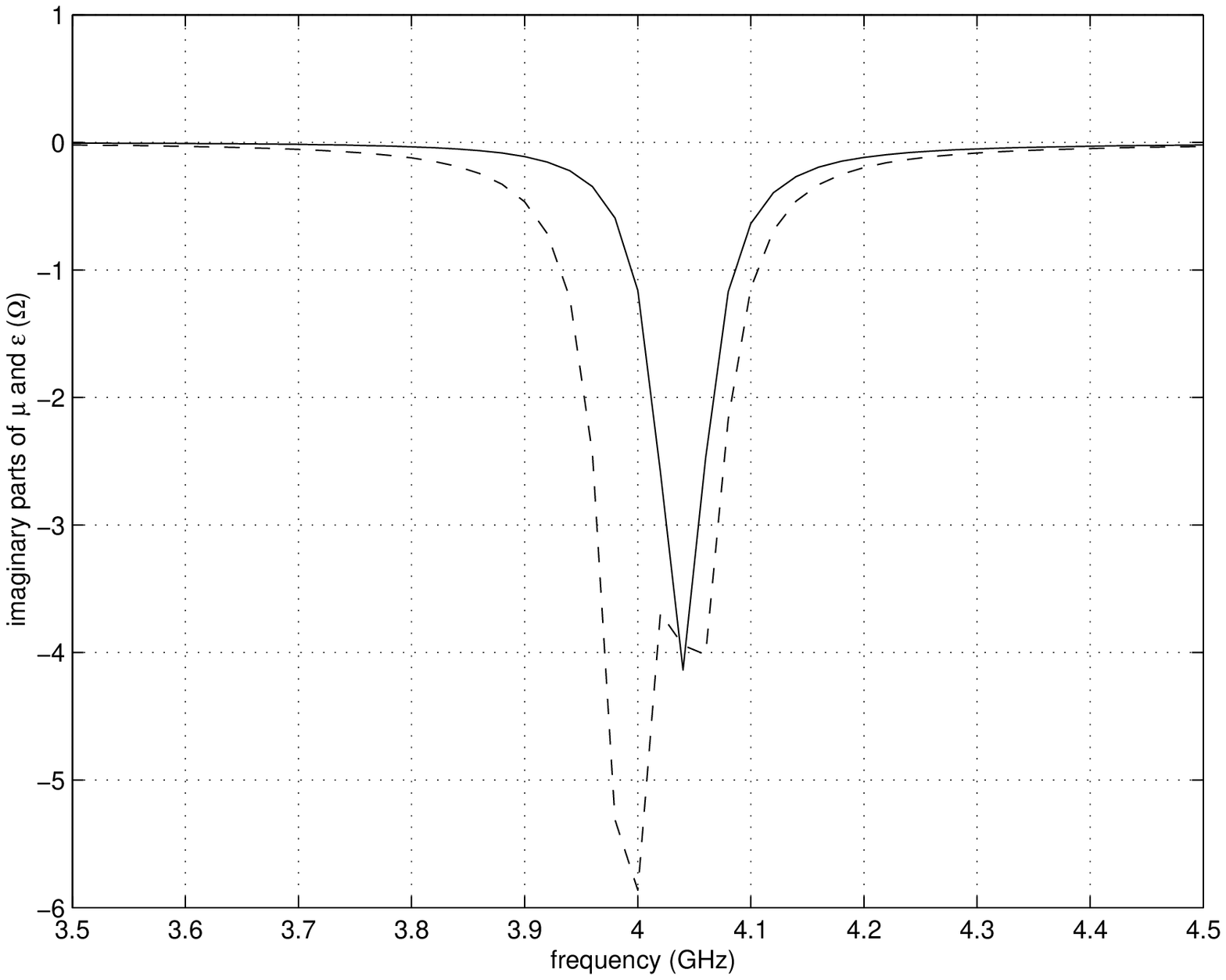, width=10cm}
\caption{Imaginary parts of the effective permeability (solid) and
permittivity (dashed) of an Omega composite  formed by cubic cells
versus frequency. The  host permittivity is $2$ and  and the length
of a unit cells is $6.5$ mm} \label{fig44}
\end{figure}

\begin{figure}
\centering \epsfig{file=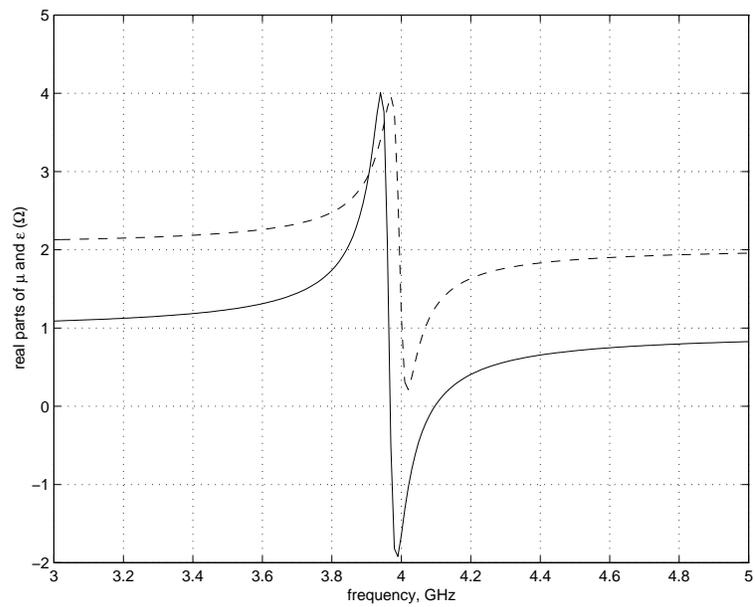, width=10cm} \caption{Real
part of the effective permeability (solid) and permittivity (dashed)
of a random Omega composite. The host permittivity is $2$.}
\label{random_omega}
\end{figure}

Now, consider the capacitances included in the loops splits. We can
write in accordance with Fig. \ref{fig2} \begin{equation}
C_1=C_{s1}+{C_{mut}\over 2},\quad C_2=C_{s2}+{C_{mut}\over 2}
\l{ccc}\end{equation} For $C_s$ we have the simple equation of a
conventional (parallel-plate) capacitor \begin{equation} C_s={\pi
r^2_0\epsilon\E0\over \delta} \l{slit}\end{equation} if the split
width $\delta$ is much smaller than the wire cross section diameter.
Otherwise, if $\delta\gg r_0$ one has $C_{s1,2}\ll C_{mut}$ and the
capacitance of the split can be ignored. We do not study the cases
when $r_0\approx \delta$.

Next, we calculate $C_{mut}$. The capacitance per unit length of two
parallel wires of diameter $2r_0$ with the distance between the wire
axes $d=a_2-a_1$ denoted below as $C_0$ is known \cite{Combes}:
\begin{equation} C_0={\pi\epsilon \E0\over {\rm arccosh}\left({d\over
2r_0}\right)} \l{C0}\end{equation} Roughly (but successfully) we can
consider $C_{mut}$ as the capacitance between two equivalent parallel
straight wires of length $\pi a\equiv\pi (a_1+a_2)/2$ with distance
$d$ between their axes. However, we would obtain a bad result if we
simply multiply $C_0$ by $\pi a$ in order to find $C_{mut}$. We
should take into account the non-uniformity of the charge
distribution along these equivalent wires and then write $
C_{mut}=C_0P$ (where the effective length of two parallel charged
wires $P$ is smaller than $\pi a$ due to this non-uniformity).

We use the King theory of loop antennas \cite{LoLee} in order to find
effective length $P$. For it we write the Fourier expansion for the
current induced in the reference loop by a magnetic field (with
respect to the azimuth angle $\phi$): \begin{equation}
I(\phi)=I^{(0)}+I^{(1)} \cos \phi +I^{(2)} \sin \phi+\dots
\l{tok}\end{equation} It corresponds to the following expansion for
the induced charges per unit angle: \begin{equation}
\tau(\phi)=\tau_1 \sin \phi +\tau_2\cos \phi+ \dots
\l{tau}\end{equation} In the King theory it was proven that (for the
case of small loop antenna ($ka_{1,2}\ll 1$) with a finite impedance
of the split)  the omitted terms in expansions \r{tok} and \r{tau}
are of the next order of smallness compared to the remained terms
\cite{LoLee}. The uniform term $I^{(0)}$ represents the
magneto-dipole mode of the induced current and has nothing to do with
induced charges. Two other terms in \r{tok} are responsible for the
electric polarization of the loop and are related with the charge
distribution. We can rewrite \r{tau} as: \begin{equation}
\tau(\phi)=\tau_{max}\sin(\phi+\phi_0) \l{tau1}\end{equation} where
$\tau_{max}\equiv \sqrt{\tau^2_1+ \tau^2_2}$. So, at it was already
mentioned, half of the  loop ($\phi_0<\phi<\pi+\phi_0$) contains
positive charge $Q$, and the other half ($-\pi+\phi_0<\phi<\phi_0$)
contains negative charge $-Q$. The distribution of these charges is
sinusoidal. The induced charges in the other loop have the opposite
sign but the same sinusoidal distribution (see also Fig. \ref{fig2}).

We obtain that the distribution of the charge along two effective
equivalent parallel wires of length $\pi a$ is sinusoidal with zeros
at the ends of these wires. It leads to the result $P=2a$ which gives
together with \r{C0} the approximate relation for the capacitance
$C_{mut}$: \begin{equation} C_{mut}= {2 \pi\epsilon \E0 a\over {\rm
arccosh}\left({d\over 2r_0}\right)} \l{cap}\end{equation} If the loop
is excited by $y-$polarized electric field, the formula of the
current distribution \r{tok} keeps valid \cite{LoLee}. Coefficients
$I^{(0,1,2)}$ change in this case, but the charges $+Q$ and $-Q$
remain sinusoidally distributed on the two halves of each the loop.
Therefore the formula \r{cap} keeps valid, too.

\end{document}